\documentclass[aps,pra,twocolumn,amsmath,amssymb,showpacs,nofootinbib]{revtex4}

\newcommand{\bra}[1]{\langle#1|}
\newcommand{\ket}[1]{|#1\rangle}
\newcommand{\op}[2]{\hat{\textbf{#1}}_{#2}}
\newcommand{\dagop}[2]{\hat{\textbf{#1}}_{#2}^\dag}
\newcommand{\etal}{\mbox{\emph{et\ al.}} }
\newcommand{\ie}{\mbox{\emph{i.e.}} }
\usepackage[dvips]{graphicx}
\usepackage{mathrsfs}

\begin{document}

\title{Optimal photons for quantum information processing}

\author{Peter P. Rohde}
\email[]{rohde@physics.uq.edu.au}
\homepage{http://www.physics.uq.edu.au/people/rohde/}
\author{Timothy C. Ralph}
\affiliation{Centre for Quantum Computer Technology, Department of Physics\\ University of Queensland, Brisbane, QLD 4072, Australia}
\author{Michael A. Nielsen}
\affiliation{Department of Physics, University of Queensland, Brisbane, QLD 4072, Australia}

\date{\today}

\begin{abstract}
Photonic quantum information processing schemes, such as linear optics quantum computing, and other experiments relying on single-photon interference, inherently require complete photon indistinguishability to enable the desired photonic interactions to take place. Mode-mismatch is the dominant cause of photon distinguishability in optical circuits. Here we study the effects of photon wave-packet shape on tolerance against the effects of mode-mismatch in linear optical circuits, and show that Gaussian distributed photons with large bandwidth are optimal. The result is general and holds for arbitrary linear optical circuits, including ones which allow for post-selection and classical feed-forward.
\end{abstract}

\pacs{03.67.Lx,42.50.-p}

\maketitle

\section{Introduction}
The interference of single photons plays a central role in linear optics quantum computing \cite{bib:KLM01} and many other quantum optics experiments. One of the major obstacles facing experimentalists is mode-mismatch, whereby photon indistinguishability is compromised within a circuit, resulting in loss of quantum gate fidelity. The effects of input distinguishability, and, more generally, mode-mismatch, in linear optical circuits has been studied in a number of situations \cite{bib:RohdeRalph05,bib:RohdePryde04,bib:URen03,bib:Kiraz04}.

The tolerance of optical circuits against the effects of mode-mismatch is highly dependent upon the shape of the interacting photons' wave-packets. In this paper we study this effect and derive conditions which maximize the tolerance of optical circuits against the effects of mode-mismatch. We begin by considering the most trivial case of two photons interacting on a beamsplitter. We then generalize our findings and show that for arbitrary linear optical circuits, including ones which incorporate post-selection and classical feed-forward, tolerance against the effects of mode-mismatch is maximized when utilizing Gaussian shaped photons, which are as broad as possible in the degrees of freedom in which mode-mismatch is introduced.

Understanding the influence wave-packet shape has on the effects of mode-mismatch is important from a practical perspective, where experimentalists must choose the most appropriate photon engineering techniques. We provide a discussion of such techniques, in the context of our findings, in Section \ref{sec:discussion}.

\section{Proof that Gaussian is optimal for two photons interacting on a beamsplitter}
We begin by considering the most basic linear optics network: two photon interference on a beamsplitter. A beamsplitter with reflectivity $\eta$ is described by the Heisenberg equations of motion
\begin{eqnarray}
\op{a}{\mathrm{out}}&=&\sqrt{\eta}\op{a}{}+\sqrt{1-\eta}\op{b}{}\nonumber\\
\op{b}{\mathrm{out}}&=&\sqrt{\eta}\op{b}{}-\sqrt{1-\eta}\op{a}{}
\end{eqnarray}
where we assume the phase-asymmetric beamsplitter convention. $\op{a}{}$ and $\op{b}{}$ are the usual photon annihilation operators for the two spatial modes.

If we consider an $\eta=0.5$ beamsplitter with a single photon incident at each input (\ie $\ket{\psi_\mathrm{in}}=\ket{1}_\mathrm{a}\ket{1}_\mathrm{b}$), the output state it given by
\begin{equation}
\ket{\psi_\mathrm{out}}=\frac{1}{\sqrt{2}}(\ket{2}_\mathrm{a}\ket{0}_\mathrm{b}-\ket{0}_\mathrm{a}\ket{2}_\mathrm{b})
\end{equation}
Thus we see complete suppression of single photon terms as a result of quantum interference.

Next we consider the non-ideal case, where mode-mismatch is present. We model mode-mismatch in the same manner described by \cite{bib:RohdePryde04}. First we associate a wave-function $\psi(x)$, with input photons, where $x$ is some photonic degree of freedom. Note that modeling mode-mismatch in a single degree of freedom is sufficient to characterize arbitrary mode-matching effects, due to the inherent indistinguishability of the effects of mode-mismatch in different degrees of freedom. Thus, photons are represented as
\begin{equation}
\ket{\psi_\mathrm{photon}}=\int_{-\infty}^{\infty}\psi(x)\dagop{a}{}(x)\,\mathrm{d}x\,\ket{0}
\end{equation}
where $\dagop{a}{}(x)$ is the photonic creation operator at the infinitesimal point $x$. Mode-mismatch between photons of this form is represented by displacing the photons' wave-functions. This has the effect of transforming the wave-function according to $\psi(x)\to\psi(x-\tau)$, where $\tau$ is the displacement parameter.

If we input a photon into each input of an $\eta=0.5$ beamsplitter, where there is a relative displacement between the photons, we no longer observe complete suppression of the single photon terms as before. Instead the probability of measuring a coincidence between the outputs (\ie a single photon at each output) is given by
\begin{equation}
\langle\hat{N}_\mathrm{a}\hat{N}_\mathrm{b}\rangle=\frac{1}{2}-\frac{1}{2}\Big|\int_{-\infty}^{\infty}\psi(x)^*\psi(x-\tau)\,\mathrm{d}t\Big|^2
\end{equation}
For a derivation see \cite{bib:RohdeRalph05}. This expression has the property that $0\leq\langle\hat{N}_\mathrm{a}\hat{N}_\mathrm{b}\rangle\leq0.5$, with $\langle\hat{N}_\mathrm{a}\hat{N}_\mathrm{b}\rangle=0$ if and only if $\tau=0$. This is completely equivalent to the effect observed by Hong, Ou and Mandel (HOM) \cite{bib:HOM87}. The vanishing of $\langle\hat{N}_\mathrm{a}\hat{N}_\mathrm{b}\rangle$ at $\tau=0$ is widely referred to as the `HOM-dip'.

The behavior of $\langle\hat{N}_\mathrm{a}\hat{N}_\mathrm{b}\rangle$ for non-zero $\tau$ is highly dependent on the form of $\psi(x)$. We are therefore motivated to ask what form of $\psi(x)$ minimizes the effect of $\tau$ on $\langle\hat{N}_\mathrm{a}\hat{N}_\mathrm{b}\rangle$, \ie which maximizes the system's tolerance against the effects of mode-mismatch.

We assume that mode-mismatch is some unknown small deviation from the ideal case. This assumption is justified, since if the displacement parameter were known and large it could be corrected for.

We ask what form of $\psi(x)$ minimizes the curvature of the function $\langle\hat{N}_\mathrm{a}\hat{N}_\mathrm{b}\rangle$ against $\tau$. Thus we aim to minimize the function 
\begin{eqnarray}
S&=&\frac{\partial^2}{\partial\tau^2}\Big|_{\tau=0}\langle\hat{N}_\mathrm{a}\hat{N}_{b}\rangle\nonumber\\
&\propto&-\int_{-\infty}^\infty\psi(x)^*\psi^{''}(x)\,\mathrm{d}x
\end{eqnarray}
subject to the normalization constraint
\begin{equation}
\int_{-\infty}^{\infty}|\psi(x)|^2\,\mathrm{d}x=1
\end{equation}

This optimization has a trivial solution. Namely, if we choose any form for $\psi(x)$ and let its width approach infinity, the function $S$ will exhibit no dependence on $\tau$. Thus we can immediately establish the following criteria for optimal photon engineering: photon wave-functions should be as broad as possible in the degree of freedom in which mode-mismatch is introduced. For example, in the presence of temporal mode-mismatch we should make photons as temporally broad as possible. We note that this leads to a trade-off between temporal mode-mismatch and clock-speed: the longer photons are, the slower a circuit can be operated. Similarly, there are limitations in how broad photons can be in other degrees of freedom, for example spatially. This result is very intuitive, since we expect that the closer a wave-function is to being invariant under translations (\ie broader), the more stable circuit operation will be against such translations.   

Next we impose the additional constraint that $\psi(x)$ must have fixed variance,
\begin{equation}
\int_{-\infty}^{\infty}x^2|\psi(x)|^2\,\mathrm{d}x=\Delta x
\end{equation}
We have assumed the mean, $\int_{-\infty}^{\infty}x|\psi(x)|^2$, vanishes. This is allowed because our analysis is invariant under global translations in the $x$ coordinate. Fixing the variance allows us to avoid the trivial solution, and also gives us a means by which to compare different functions. Thus, the optimizing function $\psi(x)$ corresponds to the function, which, for a given bandwidth, maximizes tolerance against the effects of mode-mismatch.

\begin{figure}[!htb]
\includegraphics[width=\columnwidth]{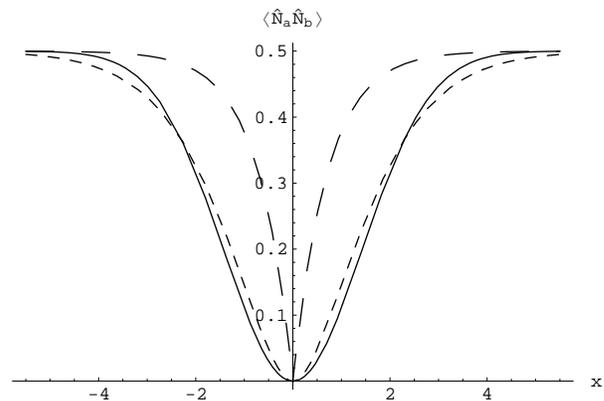}
\caption{Hong-Ou-Mandel dip for Gaussian (solid), Lorentzian (broadly dashed) and double-sided Lorentzian (finely dashed) wave-functions, where $\Delta x=1$.} \label{fig:HOM_dip}
\end{figure}

Figure \ref{fig:HOM_dip} illustrates the behavior of $\langle\hat{N}_\mathrm{a}\hat{N}_\mathrm{b}\rangle$ against $\tau$ for Lorentzian, double-sided Lorentzian and Gaussian wave-functions\footnote{Lorentzian: $\psi(x)=\sqrt[4]{\frac{2}{\pi^2}}\frac{1}{1+\sqrt{2}ix}$, double-sided Lorentzian: $\psi(x)=\sqrt[4]{\frac{2}{4\pi^2}}\frac{2}{1+2x^2}$, Gaussian: $\psi(x)=\sqrt[4]{\frac{2}{\pi}}e^{-x^2}$.}, where $\Delta x=1$.
It is immediately obvious from Figure \ref{fig:HOM_dip} that wave-packet shape plays a critical role in tolerance against the effects of mode-mismatch. Most notably, for Lorentzian distributed photons, which are characterized by a discontinuity, the coincidence rate increases far more rapidly than either the Gaussian or double-sided Lorentzian cases, both of which are smooth functions.

We can reformulate the optimization problem in a more familiar quantum mechanical description as
\begin{equation}
S\propto\langle\psi|\hat{p}^2|\psi\rangle
\end{equation}
subject to the constraints
\begin{equation}
\langle\psi|\psi\rangle=1
\end{equation}
and
\begin{equation}
\langle\psi|\hat{x}^2|\psi\rangle=\Delta x
\end{equation}
where $\hat{x}$ and $\hat{p}$ can be considered the usual position and momentum operators (but could equally correspond to any Fourier pair). The position-momentum Heisenberg uncertainty relation is given by
\begin{equation}
\langle(\Delta\hat{x})^2\rangle\langle(\Delta\hat{p})^2\rangle\geq\frac{\hbar}{4}
\end{equation}
Upon applying the constraint $\langle(\Delta\hat{x})^2\rangle=\langle\hat{x}^2\rangle=\Delta x$, this reduces to
\begin{equation}
\langle(\Delta\hat{p})^2\rangle\geq\frac{\hbar}{4\Delta x}
\end{equation}

It is known that we obtain equality for Gaussian $\psi(x)$, \ie a Gaussian state is a minimum uncertainty state \cite{bib:Sakurai94}. However $\langle(\Delta\hat{p})^2\rangle=\langle\hat{p}^2\rangle\propto S$. Thus, for Gaussian $\psi(x)$, $S$ is minimized, as required.

\section{General proof for arbitrary linear optics network}
\begin{figure}
\includegraphics[width=0.7\columnwidth]{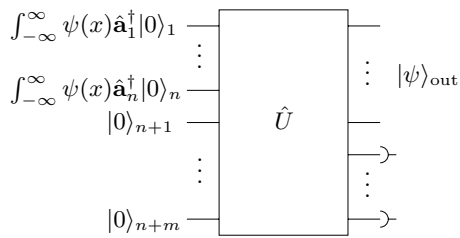}
\caption{An arbitrary linear optics network consisting of $n$ single photon inputs, all characterized by wave-function $\psi(x)$, and $m$ vacuum inputs. The detectors following some of the outputs facilitate post-selection.} \label{fig:lo_circuit}
\end{figure}

We now consider an arbitrary linear optics network of the form shown in Figure \ref{fig:lo_circuit}.
The input state, according to this model, is of the form
\begin{eqnarray}
\ket{\psi_\mathrm{in}}&=&\int_{-\infty}^{\infty}\psi(x)\dagop{a}{1}(x)\,\mathrm{d}x\int_{-\infty}^{\infty}\psi(x)\dagop{a}{2}(x)\,\mathrm{d}x\dots\nonumber\\
&&\times\int_{-\infty}^{\infty}\psi(x)\dagop{a}{n}(x)\,\mathrm{d}x\,\ket{\mathbf{0}}\nonumber\\
&=&\prod_{j=1}^{n}\int_{-\infty}^{\infty}\psi(x)\dagop{a}{j}(x)\,\mathrm{d}x\,\ket{\mathbf{0}}
\end{eqnarray}
where $\dagop{a}{i}(x)$ are the photonic creation operators of the $i$th input. Circuits where inputs contain higher photon number terms are allowed for in this model by assuming a suitable beamsplitter network to be inside the box, followed by appropriate post-selection.

If we allow $\hat{U}$ to be an arbitrary beamsplitter network acting on the input state, the output state will be of the form
\begin{equation}
\ket{\psi_\mathrm{out}}=\sum_{i_1,i_2,\dots,i_n=1}^{n+m}\lambda_{i_1,i_2,\dots i_n}\prod_{j=1}^n\int_{-\infty}^{\infty}\psi(x)\dagop{a}{i_j}(x)\,\mathrm{d}x\,\ket{\mathbf{0}}
\end{equation}
which is simply a sum-of-paths of all possible routes the input photons ($j$) could take to reach all possible output configurations ($i_1,i_2,\dots,i_j$). The complex coefficients $\lambda$ are the amplitudes of particular paths through the circuit. These parameters are a function of the circuit and completely characterize the output state. The $\lambda$ are obtained by tracing along paths from inputs to outputs. Upon reflection from or transmission through a beamsplitter, the respective parameter gains a factor equal to the beamsplitter's reflectivity or transmissivity respectively. Upon a phase change the parameter gains a complex rotation factor.

Post-selection is accommodated for through suitable adjustment of the $\lambda$ parameters and discarding the degrees of freedom associated with the measured modes. Classical feed-forward is accommodated for by recognizing that a circuit with feed-forward can be broken down into multiple blocks of the form shown in Figure \ref{fig:lo_circuit}, where the $\lambda$ parameters in later blocks are determined by measurement outcomes from earlier blocks.

\begin{figure}
\includegraphics[width=0.4\columnwidth]{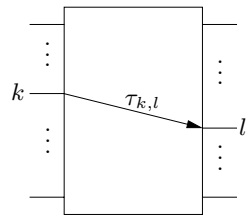}
\caption{$\tau_{k,l}$ represents the displacement introduced into the photon wave-packet when it travels from the $k$th input to the $l$th output.} \label{fig:displacement}
\end{figure}

We model mode-mismatch by introducing displacements into photon wave-packets as they travel between different inputs and outputs. We introduce the parameters $\tau_{k,l}$, which represent the cumulative displacement introduced between the $k$th input and $l$th output, shown in Figure \ref{fig:displacement}.
Thus, when a photon travels from the $k$th input to the $l$th output, its wave-function undergoes the transformation $\psi(x)\to\psi(x-\tau_{k,l})$. We have assumed that input photons are all initially indistinguishable. There is no loss of generality in making this assumption since we can take the displacement parameters describing input distinguishability to be implicitly incorporated into the box.

The general form of the output state in the presence of mode-mismatch is given by
\begin{eqnarray}
\ket{\psi_\mathrm{out}^{'}}=&&\sum_{i_1,i_2,\dots,i_n=1}^{n+m}\lambda_{i_1,i_2,\dots i_n}\nonumber\\
&&\times\prod_{j=1}^n\int_{-\infty}^{\infty}\psi(x-\tau_{j,i_j})\dagop{a}{i_j}(x)\,\mathrm{d}x\,\ket{\mathbf{0}}
\end{eqnarray}

We define the \emph{fidelity} as the overlap between the ideal and non-ideal output states,
\begin{widetext}
\begin{eqnarray}
F&=&|\langle\psi_\mathrm{out}|\psi_\mathrm{out}^{'}\rangle|^2\nonumber\\
&=&\bigg|\bra{\mathbf{0}}\Big(\sum_{i_1,i_2,\dots,i_n=1}^{n+m}\lambda_{i_1,i_2,\dots i_n}^*\prod_{j=1}^n\int_{-\infty}^{\infty}\psi(x)^*\op{a}{i_j}(x)\,\mathrm{d}x\Big)\Big(\sum_{i_1,i_2,\dots,i_n=1}^{n+m}\lambda_{i_1,i_2,\dots i_n}\prod_{j=1}^n\int_{-\infty}^{\infty}\psi(x-\tau_{j,i_j})\dagop{a}{i_j}(x)\,\mathrm{d}x\Big)\ket{\mathbf{0}}\bigg|^2\nonumber\\
&=&\bigg|\sum_{i_1,i_2,\dots,i_n=1}^{n+m}\sum_{i_1^{'},i_2^{'},\dots,i_n^{'}=1}^{n+m}\beta_{i_1,i_2,\dots i_n}^*\beta_{i_1^{'},i_2^{'},\dots i_n^{'}}\prod_{j=1}^{n}\int_{-\infty}^{\infty}\psi(x)^*\psi(x-\tau_{i_j,i_j^{'}})\,\mathrm{d}x\bigg|^2
\end{eqnarray}
\end{widetext}
where the parameters $\beta$ have been introduced to allow for the different combinations in which terms from the left-hand product can act on terms from the right-hand product.

We apply the same criteria as for the HOM case, and attempt to find the form of $\psi(x)$ such that the curvature of the fidelity function is minimized, subject to the same normalization and variance constraints as before. Thus, we wish to minimize $\frac{\partial^2F}{\partial\tau_{m,n}^2}\Big|_{\tau_{m,n}=0}$ for any given $\{m,n\}$. The result is of the form
\begin{equation}
\frac{\partial^2F}{\partial\tau_{m,n}^2}\Big|_{\tau_{m,n}=0}\propto\int_{-\infty}^{\infty}\psi(x)^*\psi^{''}(x)\,\mathrm{d}x
\end{equation}
which is the same minimization as previously. The proportionality factor is circuit dependent. Therefore, for an arbitrary linear optics network, the fidelity of the output state will be most resilient against the effects of mode-mismatch when input photons have a Gaussian profile in the degree of freedom in which the mode-mismatch is introduced.

\section{Discussion} \label{sec:discussion}
We now discuss the potential of various photon sources for quantum information processing in light of our results. For the most part we assume cavity based sources producing beams with Gaussian spatial profiles and so concentrate on their temporal profiles, although we note that this is not necessarily the case for all experimental examples cited.

\subsection{Intra-cavity spontaneous photon emission}
When a `fast' single photon emitter is placed in a `slow' optical cavity we observe photons with an approximately Lorentzian frequency wave-function,
\begin{equation}
\psi(\omega)=\frac{\kappa}{\pi}\frac{1}{\kappa+i\omega}
\end{equation}
where $\omega$ is frequency and $\kappa$ is the cavity bandwidth. Here we are assuming that $\gamma\gg\kappa$, where $\gamma$ is the spontaneous emission lifetime of the emitter. Examples of such sources include quantum dot \cite{bib:Moreau01,bib:Santori02} and fluorescence \cite{bib:Brunel99,bib:Lounis00} based sources. Typically such sources exhibit some inhomogenous broadening of the photon emission process, referred to as \emph{time-jitter}. This results in a mixing effect whereby photons are characterized by a mixture of temporally displaced Lorentzian wave-functions. To minimize this effect the time-uncertainty of photon emission must be kept small compared to the decay time of the cavity. In the presence of time-jitter the state can be expressed in the form
\begin{eqnarray}
\hat\rho&=&\int_{-\infty}^{\infty}f(\tau)\nonumber\\
&&\times\int_{-\infty}^{\infty}\int_{-\infty}^{\infty}\mathrm{e}^{\mathrm{i}(\omega-\omega^{'})\tau}\psi(\omega)\psi^*(\omega^{'})\ket{\omega}\bra{\omega^{'}}\,\mathrm{d}\omega\,\mathrm{d}\omega^{'}\,\mathrm{d}\tau\nonumber\\
\end{eqnarray}
where $f(\tau)$ is determined by the time-jitter and characterizes the mixture.

The effect of time-jitter on the Knill controlled-sign gate \cite{bib:Knill02} has been examined by Kiraz \etal \cite{bib:Kiraz04}. It was found that to achieve gate fidelity of 99\%, time-jitter must be kept below 0.3\% of the inverse bandwidth. This is in stark contrast to an analysis of the simplified KLM CNOT gate \cite{bib:Ralph01} indicating that, for Gaussian photons, gate fidelity of 99\% requires temporal synchronization to within 10\% of the inverse bandwidth \cite{bib:RohdeRalph05}. This relative intolerance against time-jitter can be attributed to the temporal discontinuity inherent in the Lorenztian function, and is the same reason we observe rapid falloff in the HOM dip for the Lorentzian case in Figure \ref{fig:HOM_dip}.

We expect that photon sources based on spontaneous emission, and other sources which produce Lorentzian photons, will not be well suited to quantum information processing applications, unless suitable filtering or other shaping techniques are first applied. Most importantly, such techniques would have to eliminate the temporal discontinuity inherent in the Lorentzian, which is the primary culprit in loss of gate fidelity in the presence of temporal mode-mismatch. In their favor, however, such sources allow production of single photons on demand, which is very desirable for quantum information processing applications.

\subsection{Non-degenerate parametric down-conversion}
Non-degenerate parametric down-conversion is widely used in quantum optics experiments for the production of heralded single photons. The down-conversion process probabilistically produces entangled photon pairs in distinct spatial modes. The output from a down-converter in an optical cavity can be expressed in the form
\begin{eqnarray}
\ket{\psi_\mathrm{out}}&=&\ket{0}_\mathrm{a}\ket{0}_\mathrm{b}+\int_{-\infty}^{\infty}\frac{2\chi\kappa}{\kappa^2+\omega^2}(\ket{0_\omega1_{-\omega}}_\mathrm{a}\ket{1_\omega0_{-\omega}}_\mathrm{b}\nonumber\\
&&+\ket{1_\omega0_{-\omega}}_\mathrm{a}\ket{0_\omega1_{-\omega}}_\mathrm{b})\,\mathrm{d}\omega
\end{eqnarray}
where $a$ and $b$ denote the spatial modes, $\chi$ is related to the conversion efficiency of the down-conversion process, and $\kappa$ to the bandwidth. It has been assumed that conversion efficiency is very weak, such that $\chi\ll\kappa$, which justifies neglecting higher-order photon number terms. The vacuum terms indicate that the down-conversion process does not always produce photon pairs. In fact, down-conversion fails the vast majority of the time.

If conditioning upon detection of a photon in one of the modes is performed, then upon success of the conditioning process there is high probability that a single photon is present in the other mode. The wave-packets of photons produced through conditioned down-conversion are characterized by a double-sided Lorentzian wave-function,
\begin{equation}
\psi(\omega)=\sqrt{\frac{\kappa}{2\pi\chi^2}}\frac{2\chi\kappa}{\kappa^2+\omega^2}
\end{equation}
where it has been assumed that the intrinsic response time of the conditioning detector $\tau_{det}$, obeys $1/\tau_{det}\gg\kappa$. The frequency distribution of down-conversion sources has been extensively studied \cite{bib:Rubin94,bib:Ou97,bib:Grice01,bib:URen03,bib:URenMukamel03}.

\begin{figure}[!htb]
\includegraphics[width=0.7\columnwidth]{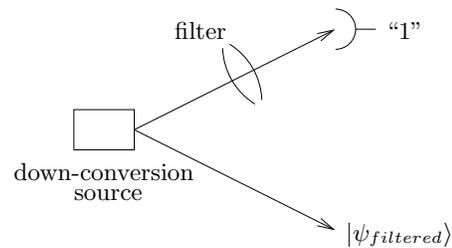}
\caption{Conditional production of filtered photons through parametric down-conversion. The down-converter produces an entangled pair of photons into two distinct spatial modes. One of the modes is filtered and a photo-detector conditions upon the detection of a single photon following the filter. When post-selection succeeds, a photon will be present in the other path, whose wave-function will reflect that of the detected filtered photon.} \label{fig:dc_source}
\end{figure}

By applying filtering to the conditioned mode, one can perform non-local pulse shaping \cite{bib:Aichele02,bib:Bellini03}. An example of such a scheme is shown in Figure \ref{fig:dc_source}. 
It has been shown by Ou \cite{bib:Ou97} and Aichele \etal \cite{bib:Aichele02} that to obtain high purity of the post-selected state, narrowband filtering must be applied to the conditioned mode. Non-local pulse shaping is not limited to the temporal domain. In principle, shaping can be performed in any degree of freedom in which the photon pair are entangled.

On the one hand, non-degenerate parametric down-conversion is quite suitable for quantum information processing applications due to its `raw' symmetric profile and the ability for non-local pulse shaping techniques to allow for approximately Gaussian (\ie optimal) photons to be engineered. These properties have made spontaneous down-conversion the system of choice for in-principle demonstrations in which efficiency is not an issue. However, although photons are heralded, scalability would require good quantum optical memories. While in-principle demonstrations have been performed \cite{bib:Pittman02}, efficiencies are currently too low to be practical. 

\subsection{Cavity QED pump pulse manipulation}
Perhaps the best solution is to combine a coherent excitation with single emitter technology. This can be achieved by using a Raman process to pump a single emitter in a high-Q cavity.

Keller \etal \cite{bib:Keller04} demonstrated that, through manipulation of the pump pulse, the temporal wave-function of photons emitted from Raman pumped single ions trapped in a cavity can be readily manipulated. Experimental results using a single trapped $^{40}\mathrm{Ca}^+$ ion, demonstrated the production of Gaussian, rectangular and double-peaked pulse formations.

\section{Conclusion}
We considered the influence of photon wave-packet shape on the effects of mode-mismatch in linear optical circuits, from which we established two criteria which optimize circuit tolerance against such effects. Firstly, photons should be as broad as possible in the degrees of freedom in which mode-mismatch is likely to be introduced. Secondly, for a given bandwidth, photons with Gaussian profile are optimal. Our findings are completely general and hold for arbitrary linear optics circuits (\ie beamsplitter networks). This includes ones which incorporate post-selection and classical feed-forward, making the findings applicable to linear optics quantum computing circuits.

We considered various photon production techniques, discussing their advantages and limitations in producing photons suitable for quantum information processing applications. Of these, some are inherently more suitable than others, within the context of our established criteria.

\begin{acknowledgments}
We thank Etera Levine and Chris Foster for helpful discussions. This work was supported by the Australian Research Council.
\end{acknowledgments}

\bibliography{paper.bib}

\end{document}